\documentclass[11pt]{article}

\usepackage{amsthm}

\newtheorem{theorem}{Theorem}[section]
\newtheorem{lemma}[theorem]{Lemma}

\newtheorem{corollary}[theorem]{Corollary}

\newenvironment{definition}[1][Definition]{\begin{trivlist}
\item[\hskip \labelsep {\bfseries #1}]}{\end{trivlist}}

\newenvironment{question}[1][Question]{\begin{trivlist}
\item[\hskip \labelsep {\bfseries #1}]}{\end{trivlist}}

\newcommand{\can}{\{0,1\}^\infty}
\newcommand{\nat}{\mathbb{N}}

\newcommand{\toinf}{\rightarrow \infty}
\newcommand{\eref}[1]{(\ref{#1})}
\newcommand{\pint}{\mathbb{Z}^+}

\usepackage{verbatim}
\usepackage{amssymb}
\usepackage{amsfonts}
\usepackage{clrscode}
\usepackage{amsmath}
\usepackage[left=2.8cm,right=2.8cm,top=2.4cm,bottom=2.4cm,nohead,foot=1cm]{geometry}

\begin{document}



\begin{center}
{\Large \ \ \ \ \ \ \ \ \ \ \ \ \ \ \ \ \ \ Pushdown Dimension }
\newline
\newline
{\large David Doty \footnote{\scriptsize Department of Computer Science, Department of Bioinformatics and Computational Biology, Iowa State University, Ames, IA, 50011, USA. {\bf ddoty $<$at$>$ iastate $<$dot$>$ edu}. This work was supported in part by National Science Foundation IGERT grant DGE-9972653.} \ Jared Nichols \footnote{\scriptsize Department of Computer Science, Iowa State University, Ames, IA, 50011, USA. This work was supported in part by National Science Foundation Grants 9988483 and 0344187.}}
\end{center}

\begin{abstract}
This paper develops the theory of pushdown dimension and explores its relationship with finite-state dimension. Pushdown dimension is trivially bounded above by finite-state dimension for all sequences, since a pushdown gambler can simulate any finite-state gambler. We show that for every rational $0 < d < 1$, there exists a sequence with finite-state dimension $d$ whose pushdown dimension is at most $\frac{1}{2} d$. This establishes a quantitative analogue of the well-known fact that pushdown automata decide strictly more languages than finite automata. 
\end{abstract}

\section{Introduction}
The dimension of a set of points was first explored by Hausdorff \cite{hausdorff,hausdorff-book}, who showed that there exist sets of points with fractional dimension, now termed \emph{fractals}. Infinite sequences drawn from a finite alphabet can be viewed as points on the unit interval. Lutz \cite{lutz-hausdorff} showed that the Hausdorff dimension of a set of infinite sequences could be characterized by the rate at which money could be taken away from a gambler that is trying to make unbounded money by betting on all the sequences in the set. In other words, the higher the dimension of a set, the more random and unpredictable are its elements, and so the more difficult it is to make money betting on its elements. (A precise definition follows in later sections).

Though all singleton sets of sequences -- i.e. all individual points -- have Hausdorff dimension 0, by restricting the computational power of the gambler, individual sequences can be assigned a non-zero dimension. Resource-bounded dimension (constructive dimension \cite{lutz-dim-indv}, pushdown dimension \cite{nichols}, finite-state dimension \cite{fsd}, etc.) is a measure of the density of information or randomness in a sequence as it appears to a gambler whose computational power is limited by the resource bound. Accordingly, the finite-state dimension of a sequence \cite{fsd} is the degree to which the sequence appears random to a finite automaton, and the pushdown dimension of  a sequence \cite{nichols} is the degree to which the sequence appears random to a pushdown automaton.

A finite-state gambler is a finite automaton that bets money on the next character according to its current state. A pushdown gambler is a finite-state gambler augmented with an infinite stack memory, and it is allowed to vary its state transition and its bet at each state depending on the character appearing at the top of the stack. Since any finite-state gambler can be simulated exactly by a pushdown gambler that makes no use of its stack, pushdown gamblers are at least as powerful as finite-state gamblers, and hence $\func{dim}_{PD}(S) \leq \func{dim}_{FS}(S)$ for all sequences $S$.

We show that there exist sequences with a pushdown dimension strictly less than their finite-state dimension. Specifically, for every rational $0 < d < 1$, there exists a sequence $S$ with $\func{dim}_{FS}(S)=d$ such that $\func{dim}_{PD}(S) \leq \frac{1}{2} \func{dim}_{FS}(S)$. It is well known \cite{kozen} that the class of languages decided by a pushdown automaton is a strict superset of the class of languages recognized by a finite automaton. The result presented here thus gives a quantitative, information-theoretic estimate of the extra power of a pushdown automaton over a finite automaton.

Weighted finite automata, of which finite-state gamblers are a special case, have also been studied in other contexts \cite{kuich-book,staiger-weighted}.

Section \ref{prelim} establishes preliminaries and notation, section \ref{dimension} defines dimension, finite-state dimension, and pushdown dimension, section \ref{fs-v-pd} establishes the separation of finite-state and pushdown dimension, and section \ref{conclusion} concludes and states open questions. The appendix contains proofs.

\section{Preliminaries}
\label{prelim}

We write $\mathbb{Q}$ for the set of all rational numbers, $\mathbb{Z}$ for the set of all integers, $\nat$ for the set of all natural numbers, $\mathbb{Z}^+$ for the set of all positive integers. Let $\log r = \log_2 r$.

Let $\Sigma$ be a finite alphabet of characters. $\Sigma^*$ is the set of all finite strings drawn from $\Sigma$. The length of a string $w \in \Sigma^*$ is denoted by $|w|$. $\lambda$ denotes the empty string. For $l \in \nat$, $\Sigma^l$ denotes the set of all strings $w \in \Sigma^*$ such that $|w| = l$. $\overline{w}$ denotes the reverse of $w$. For $w,y \in \Sigma^*$, $wy$ denotes the concatenation of $w$ and $y$. For $i \geq 1$, $w^i$ denotes the string $\underbrace{w w \ldots w}_{i \text{ times}}$.

$\Sigma^\infty$ is the set of all infinite sequences drawn from $\Sigma$. For $S \in \Sigma^\infty$ or $\Sigma^*$ and $i,j \in \nat$, we write $S[i]$ to denote the $i$'th character of $S$, with $S[1]$ being the leftmost character, and we write $S[i \twodots j]$ to denote the substring consisting of the $i$'th through $j$'th characters of $S$, with $S[i \twodots j] = \lambda$ if $i > j$. We write $S_n$ to denote $S[1 \twodots n]$, the $n$'th prefix of $S$. If $n < 1, S_n = \lambda$. For $S \in \Sigma^\infty$, we write $S[n \twodots]$ to denote $S$ without its first $n-1$ characters; i.e. $S_{n-1} S[n \twodots] = S$.

Let $l \in \pint$, $w \in \Sigma^l$ and $S \in \Sigma^\infty$. We write $\#(w,S_n)$ to denote the number of times $w$ appears as a substring of $S_n$. Let the \emph{frequency of $w$ in $S_n$} be defined
$
freq(w,S_n) \triangleq \frac{\#(w,S_n)}{n-l+1}.
$
Let the \emph{frequency of $w$ in $S$} be defined
$
freq(w,S) \triangleq \lim\limits_{n \toinf} freq(w,S_n) = \lim\limits_{n \toinf} \frac{\#(w,S_n)}{n},
$
when this limit exists. 

We state the following obvious lemma without proof, which states that adding a finite prefix to a sequence cannot alter the limiting frequency of any substring.

\begin{lemma}
Let $S \in \Sigma^\infty$ and $w,u \in \Sigma^*$. Then, if $freq(w,S)$ is defined,
$
freq(w,S) = freq(w,u S).
$
\label{lem-prefix}
\end{lemma}

A sequence $S \in \Sigma^\infty$ is \emph{normal} if, for every $w \in \Sigma^*$, $freq(w,S) = |\Sigma|^{-|w|}.$
In other words, $S$ is normal if, for every string length $l$, all strings of length $l$ occur with the same frequency.

Note that given $S_n$ and $l \leq n$, $freq(w,S_n)$ defines a probability measure on the set $\Sigma^l$. Accordingly, we can speak of the entropy of this probability distribution. Let the \emph{$l$'th normalized entropy of $S$} be denoted
$$
H_l(S) \triangleq \frac{1}{l \log |\Sigma|} \liminf_{n \toinf} \sum_{w \in \Sigma^l} freq(w,S_n) \log \frac{1}{freq(w,S_n)}.
$$
Note that this exists even if $freq(w,S)$ does not, since the $\liminf\limits_{n \toinf}$ is being used. This is the limiting entropy of the distribution of strings of length $l$ in $S$, normalized by the term $\frac{1}{l \log |\Sigma|}$ to fall between 0 and 1. Thus, the more uniformly distributed are the strings of length $l$ in $S$, the closer $H_l(S)$ is to 1.
Let the \emph{normalized entropy rate of $S$} be denoted $H(S) \triangleq \lim\limits_{l \toinf} H_l(S).$
Likewise, the closer $H(S)$ is to 1, the closer $S$ is to normal, and $H(S) = 1 \iff S$ is normal.

\section{Dimension}
\label{dimension}

\subsection{Hausdorff dimension}

The definition of dimension depends on the concept of martingales, which are strategies for gambling on character sequences, and $s$-gales (see \cite{lutz-dim-indv}), which are martingales in which some fraction of the winnings are removed after each bet.

Let $\Sigma$ be an alphabet.

\begin{definition}
($s$-gale and martingale)

\begin{enumerate}
    \item An \emph{$s$-gale} is a function $d^{(s)}: \Sigma^* \rightarrow [0,\infty)$ that satisfies
    $d^{(s)}(w) = |\Sigma|^{-s} \sum_{a \in \Sigma} d(wa).$
    \item A \emph{martingale} is a 1-gale, denoted $d(w)$.
\end{enumerate}
\end{definition}

Intuitively, a martingale is a strategy for betting in the following game. The gambler starts with some initial amount of money (usually 1), termed \emph{capital}, and it reads a sequence $S$ of characters drawn from alphabet $\Sigma$. At each step, the gambler bets some fraction of its capital on each character in $\Sigma$. All of its money must be bet. Whichever character actually comes next in the sequence, it loses all the capital it bet on the other characters, and the fraction of its capital that was bet on the character that appeared is multiplied by $|\Sigma|$. The only restriction on a martingale is that the bet it makes on character $S[i]$ of the sequence $S$ should be a deterministic function only of the string $S_{i-1}$, the characters that have appeared so far. The objective of a martingale is to make a lot of money, and it will make more money on a sequence if a larger fraction of capital is placed on the characters that actually occur in the sequence. All of the gambler's money must be bet, but it can ``bet nothing'' by betting $\frac{1}{|\Sigma|}$ of its capital on each character in $\Sigma$. In this way, no matter what character comes next, its capital will not change.

An $s$-gale is a martingale in which the amount of money the gambler bet on the character that occurred is multiplied by $|\Sigma|^s$, as opposed to simply $|\Sigma|$, after each character. Note that $s=1$ constitutes the original martingale condition. Since we will consider $0 \leq s \leq 1$, this will always constitute a reduction in capital from the capital the martingale makes. The lower the value of $s$, the faster money is taken away. Note that if a gambler's martingale for a string $w$ is $d(w)$, then its $s$-gale is given by $d^{(s)}(w) = |\Sigma|^{(s-1)|w|} d(w)$.

\begin{definition}
Let $P \subseteq \Sigma^*$. $P$ is a \emph{prefix set} if no string in $P$ is a proper prefix of any other string.
\end{definition}

Note that for any $l \in \mathbb{Z}^+$, $\Sigma^l$ is a prefix set. The following generalization of the Kraft inequality was given in \cite{lutz-dim-indv}:

\begin{lemma}
Let $s \in [0,\infty)$. If $d^{(s)}$ is an $s$-gale and $A \subseteq \{0,1\}^*$ is a prefix set, then for all $u \in \{0,1\}^*$,
$ \sum_{w \in A} 2^{-s |w|} d^{(s)}(uw) \leq d^{(s)}(u).$
\label{lem-kraft}
\end{lemma}

\begin{corollary}
Let $s \in [0,\infty)$. If $d^{(s)}$ is an $s$-gale and $A \subseteq \{0,1\}^*$ is a prefix set, then
$$ \sum_{w \in A} 2^{-s |w|} d^{(s)}(w) \leq 1.$$
\label{cor-kraft}
\end{corollary}

\begin{definition}
($s$-success)

Let $S \in \Sigma^\infty$. We say that a gambler \emph{$s$-succeeds on $S$} if the gambler's $s$-gale $d^{(s)}$ satisfies
$\limsup_{n \toinf} d^{(s)}(S_n) = \infty.$
If the gambler 1-succeeds on $S$, we say it \emph{succeeds} on $S$.
\end{definition}

So a gambler $s$-succeeds on $S$ if it makes an unbounded amount of money betting on $S$, even with money taken away at rate $s$.

\begin{definition}
(dimension of a set)

Let $\mathbb{S} \subseteq \Sigma^\infty$. Then the \emph{Hausdorff dimension} of $\mathbb{S}$ is
$$\func{dim}_H(\mathbb{S}) = \inf \left\{ s \in [0,\infty) \ \left| \ \begin{array}{c} \exists \text{ an $s$-gale $d$ that $s$-succeeds } \\ \text{ on every sequence $S \in \mathbb{S}$} \end{array} \right.  \right\}.$$
\end{definition}

\subsection{Finite-state dimension}
Finite-state dimension is defined in analogy to Hausdorff dimension, where the gamblers implementing the $s$-gales are restricted to have finite-state computational power. In order to define finite-state dimension, then, we must first define finite-state gamblers, which were first studied by Feder \cite{feder}. Finite-state gamblers can be thought of as weighted finite automata \cite{kuich-book}, where the weight on each state represents a percentage of money to bet on the next character.

Let $\Sigma$ be an alphabet. Let $\Delta_{\mathbb{Q}}(\Sigma)$ be the set of all rational probability measures over $\Sigma$.

\begin{definition}
(finite-state gambler \cite{fsd,feder})

A \emph{finite-state gambler} (FSG) is a 5-tuple $G = (Q,\Sigma,\delta,\beta,q_0)$ where
\begin{itemize}
    \item $Q$ is a finite set of \emph{states},
    \item $\Sigma$ is the finite \emph{input alphabet},
    \item $\delta: Q \times \Sigma \rightarrow Q \cup \{\bot\}$ is the \emph{transition function},
    \item $\beta: Q \rightarrow \Delta_{\mathbb{Q}}(\Sigma)$ is the \emph{betting function},
    \item $q_0 \in Q$ is the \emph{start state}.
\end{itemize}
\end{definition}

We write $FSG$ to mean the set of all finite-state gamblers.

If $\delta(q,a) = \bot$, for some $q \in Q$ and $a \in \Sigma$, then that transition is undefined. We extend $\delta$ to take strings as input with the function $\delta^*: Q \times \Sigma^* \rightarrow Q$ defined by
\begin{eqnarray*}
\delta^*(q,\lambda) &=& q, \\
\delta^*(q,wa) &=& \delta(\delta^*(q,w),a).
\end{eqnarray*}
for all $q \in Q, w \in \Sigma^*$, and $a \in \Sigma$. $\delta^*$ is then abbreviated $\delta$ and $\delta(q_0,w)$ is abbreviated $\delta(w)$. Intuitively, this allows us to identify $\delta(w)$ as ``the state $G$ is in after reading string $w$.''

\begin{definition}
(finite-state $s$-gale)

Let $G \in FSG$ be a finite-state gambler. The \emph{$s$-gale for $G$} is the function $d_G^{(s)}: \Sigma^* \rightarrow [0,\infty)$ defined by
\begin{eqnarray*}
d_G^{(s)}(\lambda) &=& 1 \\
d_G^{(s)}(wa) &=& d_G(w) \cdot \beta(\delta(w))(a) \cdot |\Sigma|^s
\end{eqnarray*}
for all $s \in [0,\infty)$, $w \in \Sigma^*$, and $a \in \Sigma$.
\end{definition}

Intuitively, the $s$-gale for an FSG $G$ is determined as follows. An FSG $G=(Q,\Sigma,\delta,\beta,q_0)$ starts in state $q_0$ with initial capital 1. Assuming that after some time $G$ has capital $c$ and is in state $q$, the bet it makes on each character $a \in \Sigma$ is given by $\beta(q)(a)$. Assuming the character $b$ appears next in the sequence, $G$ then transitions to state $\delta(q,b)$, and its capital becomes $c \cdot \beta(q)(b) \cdot |\Sigma|^s$.

Let $G=(Q,\Sigma,\delta,\beta,q_0)$ be an FSG. For $q \in Q$, let $d_{G,q}^{(s)}$ be the $s$-gale for $G$ if $G$ is started in state $q$ instead of $q_0$.

\begin{definition}
(finite-state dimension)

Let $S \in \Sigma^\infty$. The \emph{finite-state dimension of $S$} is
$$\func{dim}_{FS}(S) = \inf \left\{ s \in [0,\infty) \ \left|\ ( \exists G \in FSG) \limsup_{n \toinf} d_G^{(s)}(S_n) = \infty \right. \right\}.$$
\end{definition}

Thus, if $s > \func{dim}_{FS}(S)$, then there is a finite-state gambler $G$ that $s$-succeeds on $S$, meaning $G$ can make unlimited money betting on $S$, even if its winnings are multiplied by $|\Sigma|^{s-1}$ after every character.

Let $\Sigma \subseteq \Sigma'$, and let $S \in \Sigma^\infty$. Let $\func{dim}_{FS}^{(\Sigma')}(S)$ be the finite-state dimension of $S$ when considered as a sequence drawn from alphabet $\Sigma'$, even though it is really drawn from $\Sigma$.

\begin{lemma}
Let $S \in \Sigma^\infty$, and let $\Sigma \subseteq \Sigma'$. Then
$\func{dim}_{FS}^{(\Sigma')}(S) = \frac{\log |\Sigma|}{\log |\Sigma'|} \func{dim}_{FS}^{(\Sigma)}(S).$
\label{lem-add-chars-dim}
\end{lemma}

\subsection{Pushdown dimension}
Pushdown dimension \cite{nichols} will be defined almost exactly as finite-state dimension, but the gamblers will have an infinite stack memory, which will allow them to alter both their state transitions and their bets based on the symbol currently on top of the stack.

\begin{definition}
(pushdown gambler)

A pushdown gambler (PDG) is a 7-tuple $P = (Q,\Sigma,\Gamma,\delta,\beta,q_0,z)$
where
\begin{itemize}
    \item $Q$ is a finite set of \emph{states},
    \item $\Sigma$ is the finite \emph{input alphabet},
    \item $\Gamma$ is the finite \emph{stack alphabet},
    \item $\delta: Q \times \Gamma \times (\Sigma \cup \{ \lambda \})\rightarrow (Q \times \Gamma^*) \cup \{\bot\}$ is the \emph{transition function},
    \item $\beta: Q \times \Gamma \rightarrow \Delta_{\mathbb{Q}}(\Sigma)$ is the \emph{betting function},
    \item $q_0 \in Q$ is the \emph{start state},
    \item $z \in \Gamma$ is the \emph{stack start symbol}.
\end{itemize}
\end{definition}

We write $PDG$ to mean the set of all pushdown gamblers.

Note that the transition function $\delta$ outputs a next state and a string $w \in \Gamma^*$. The top character is always popped and replaced with $w$. If $a$ is the symbol currently on top of the stack, and $P$ needs to add a character $b$ to the top, it pushes the string $ba$. If it needs to leave the stack alone, it pushes the string $a$. If it needs to pop a character, it pushes the string $\lambda$. Note that the strings are pushed in reverse order; the last character of the string is pushed first.

Note also that the transition function $\delta$ accepts $\lambda$ as an input character in addition to elements of $\Sigma$. $P$ has the option not to read an input character and instead to alter the stack. To enforce determinism, we require at least one of the following hold for all $q \in Q$ and all $a \in \Gamma$:
\begin{itemize}
    \item $\delta(q,a,\lambda) = \bot$,
    \item $\delta(q,a,b) = \bot$ for all $b \in \Sigma$.
\end{itemize}
The determinism condition requires that the PDG can't have the choice to read 0 or 1 characters; the number of characters read is entirely a function of the state and the character at the top of the stack.

We must also handle the special case that the stack start symbol gets popped. Since this represents the bottom of the stack, we restrict $\delta$ so that $z$ cannot be removed from the bottom. We restrict $\delta$ so that, for every $q \in Q$ and $a \in \{\lambda\} \cup \Sigma$, either $\delta(q,z,a) = \bot$, or $\delta(q,z,a) = (q',vz)$, where $q' \in Q$ and $v \in \Gamma^*$.

As before, if $\delta(q,a,b)=\bot$ for some $q \in Q,a \in \Gamma$, and $b \in \{\lambda\} \cup \Sigma$, then that transition is undefined. We extend $\delta$ to the transition function
$$\delta^*: Q \times \Gamma^+ \times (\{\lambda\} \cup \Sigma) \rightarrow (Q \times \Gamma^*) \cup \{ \bot \},$$
defined for all $q \in Q, a \in \Gamma, v \in \Gamma^*$, and $b \in \Sigma$ as follows:
$$
\delta^*(q,av,b) = \left\{%
\begin{array}{ll}
    ( \delta_Q(q,a,b) , \delta_\Gamma(q,a,b) v ), & \hbox{if $\delta(q,a,b) \neq \bot$;} \\
    \bot, & \hbox{otherwise.} \\
\end{array}%
\right.
.$$
where $\delta(q,a,b) = ( \delta_Q(q,a,b) , \delta_\Gamma(q,a,b) )$. $\delta^*$ is then abbreviated as $\delta$. We then use the extended transition function
$$\delta^{**}: Q \times \Gamma^+ \times \Sigma^* \rightarrow (Q \times \Gamma^*) \cup \{ \bot \},$$
in analogy to that used with finite-state gamblers, defined for all $q \in Q, a \in \Gamma, v \in \Gamma^*, w \in \Sigma^*$, and $b \in \Sigma$ by
\begin{eqnarray*}
    \delta^{**}(q,av,\lambda)
    &=&
    \left\{%
\begin{array}{ll}
   \delta^{**}(\delta(q,av,\lambda),\lambda) , & \text{if } \delta(q,av,\lambda) \neq \bot \\
   (q,av) , & \text{otherwise} \\
\end{array}%
\right.
,\\
    \delta^{**}(q,av,wb)
&=&
    \left\{%
\begin{array}{ll}
   \delta^{**}(\delta(\delta^{**}(q,av,w),\lambda),b) , & \text{if } \delta^{**}(q,av,w) \neq \bot \text{ and } \delta(\delta^{**}(q,av,w),\lambda) \neq \bot \\
   \delta(\delta^{**}(q,av,w),b) , & \text{if } \delta^{**}(q,av,w) \neq \bot \text{ and } \delta(\delta^{**}(q,av,w),\lambda) = \bot \\
   \bot , & \text{otherwise} \\
\end{array}%
\right.
.\end{eqnarray*}

We then abbreviate $\delta^{**}$ to $\delta$, and $\delta(q_0,z,w)$ to $\delta(w)$. Informally, this allows us to use $\delta(w)$ as shorthand for ``the state and contents of the stack of the gambler $P$ after reading string $w$''.

We also extend $\beta$ for convenience to the function
$$ \beta^* : Q \times \Gamma^+ \rightarrow \Delta_{\mathbb{Q}},$$
defined for all $q \in Q$, $a \in \Gamma$, and $v \in \Gamma^*$ by
$$
\beta^*(q,av) = \beta(q,a).
$$
$\beta^*$ is then abbreviated $\beta$. $\beta^*(q,av)(b)$ means, informally, ``The amount bet on character $b$ when in state q, when the string $av$ is on the stack.'' Note that only the top character $a$ of $av$ can affect any single bet, but for the purpose of examining multiple steps of the gambler, it may be necessary to keep track of what the entire contents of the stack are, since they may change from step to step.

Pushdown $s$-gales and dimension are defined exactly as their finite-state versions.

\begin{definition}
(pushdown $s$-gale)

Let $P \in PDG$ be a pushdown gambler. The \emph{$s$-gale for $P$} is the function $d_P^{(s)}: \Sigma^* \rightarrow [0,\infty)$ defined by
\begin{eqnarray*}
d_P^{(s)}(\lambda) &=& 1, \\
d_P^{(s)}(wa) &=& d_P(w) \cdot \beta(\delta(w))(a) \cdot |\Sigma|^s.
\end{eqnarray*}
for all $s \in [0,\infty)$, $w \in \Sigma^*$, and $a \in \Sigma$.
\end{definition}

Intuitively, the $s$-gale for a PDG $P$ is determined as follows. A PDG $P=(Q,\Sigma,\Gamma,\delta,\beta,q_0,z)$ starts in state $q_0$ with initial capital 1. Assuming that after some time $P$ has capital $c$, is in state $q$, and the character on the top of the stack is $t$, the bet it makes on each character $a \in \Sigma$ is given by $\beta(q,t)(a)$. Assuming the character $b$ appears next in the sequence, if $\delta(q,t,b) = (q',w)$  $G$ then transitions to state $q'$, pops the top character off of the stack and replaces it with the string $w$, and its capital becomes $c \cdot \beta(q,s)(b) \cdot |\Sigma|^s$.

\begin{definition}
(pushdown dimension)

Let $S \in \Sigma^\infty$. The \emph{pushdown dimension of $S$} is
$$\func{dim}_{PD}(S) = \inf \left\{ s \in [0,\infty) \ \left|\ (\exists P \in PDG) \limsup_{n \toinf} d_P^{(s)}(S_n) = \infty \right. \right\}.$$
\end{definition}

Thus, if $s > \func{dim}_{PD}(S)$, then there is a pushdown gambler $P$ that $s$-succeeds on $S$, meaning $P$ can make unlimited money betting on $S$, even if its winnings are multiplied by $|\Sigma|^{s-1}$ after every character.

Pushdown gamblers are then nothing more than finite-state gamblers that make use of an unbounded stack memory, the top character of which can be used to inform the transition and betting functions. Additionally, PDG's are allowed to delay reading the next character of the input -- they read $\lambda$ from the input -- in order to alter the contents of their stack. During such a $\lambda$-transition, the gambler's capital remains unchanged.

\section{Finite-state versus pushdown dimension}
\label{fs-v-pd}

\subsection{Marker characters and finite-state dimension}
This section establishes that adding marker characters to a sequence, where the marker is not in the alphabet of the sequence, does not alter the finite-state dimension of the sequence, as long as the markers are spaced far enough apart. In other words, the addition of the markers cannot significantly hurt or help a finite-state gambler.

Recall that, for $S \in \Sigma^\infty$,
$$
H(S) \triangleq \lim\limits_{l \toinf} \frac{1}{l \log |\Sigma|} \liminf\limits_{n \toinf} \sum_{w \in \Sigma^l} freq(w,S_n) \log \frac{1}{freq(w,S_n)}.
$$
Let
$$
\widehat{H}(S) \triangleq \lim\limits_{l \toinf} \frac{1}{l \log |\Sigma|} \limsup\limits_{n \toinf} \sum_{w \in \Sigma^l} freq(w,S_n) \log \frac{1}{freq(w,S_n)}.
$$
Lempel and Ziv \cite{ziv-lempel} showed that
$
\widehat{\rho}_{FS}(S) = \widehat{H}(S),
$
where $\widehat{\rho}_{FS}(S)$ is the optimal compression ratio achievable by any finite-state compressor (see \cite{ziv-lempel} or \cite{fsd} for a more complete description). Dai, et al. \cite{fsd} showed that $\func{dim}_{FS}(S)$ is identical to a slightly modified form of $\widehat{\rho}_{FS}(S)$. A straightforward modification of the proof of Lempel and Ziv, combined with the result of Dai, et al., yields the following lemma relating finite-state dimension to entropy.
\begin{lemma}
Let $S \in \Sigma^\infty$. Then
$
\func{dim}_{FS}(S) = H(S).
$
\label{lem-dim-h-eq}
\end{lemma}

\begin{corollary}
Let $S \in \Sigma^\infty$. Then
$
\func{dim}_{FS}(S) = 1 \iff S \text{ is normal.}
$
\label{cor-dim-normal}
\end{corollary}

Let $\Sigma$ be an alphabet. Let $\Sigma_m = \Sigma \cup \{ m \}$, where $m \not\in \Sigma$ is a \emph{marker character}. Recall that $\func{dim}_{FS}^{(\Sigma_m)}(S)$ is the finite-state dimension of $S$ when considered as a sequence drawn from alphabet $\Sigma_m$, even if it is really drawn from $\Sigma \subsetneq \Sigma_m$.

\begin{lemma}
Let $S \in \Sigma^\infty$.
Let $S' \in \Sigma_m^{\infty}$ be constructed from $S$ by inserting the character $m$ immediately after the positions $i_1 < i_2 < i_3 \ldots$ in $S$ such that the function $f(j) = i_{j+1} - i_j$ is nondecreasing and unbounded. Then
$
\func{dim}_{FS}^{(\Sigma_m)}(S') = \func{dim}_{FS}^{(\Sigma_m)}(S).
$
\label{lem-dim-same}
\end{lemma}

\subsection{Bitstring characters and finite-state dimension}
\label{sec-bitstring-char}
In this section, we will interpret bitstrings of length $l$ to be characters, the alphabet of the sequence will be a subset of $\{0,1\}^l - \{1^l\}$, and the marker ``character'' will be $1^l$. An infinite binary sequence $S \in \{0,1\}^\infty$ will then be simultaneously interpreted as an infinite sequence $S \in A^\infty$, where $A \subset \{0,1\}^l$. In other words, every $l$ bits of $S$ will constitute 1 character from $A$. We interpret $\func{dim}_{FS}^{(\{0,1\})}(S)$ to be the finite-state dimension of $S$ when viewed as an infinite binary sequence, and we interpret $\func{dim}_{FS}^{(A)}(S)$ to be the finite-state dimension of $S$ when viewed as an infinite sequence drawn from $A$.

Note that this interpretation of $\func{dim}_{FS}^{(A)}(S)$ is different from the meaning of $\func{dim}_{FS}^{(A)}(S)$ when $\{0,1\} \subseteq A$ (i.e. in the sense of Lemma \ref{lem-add-chars-dim}). In the current case, the boundaries between characters actually change when moving from alphabet $\{0,1\}$ to alphabet $A$, in that a string of $l$ bits is required to constitute one character of $A$. In the former case, for $\Sigma \subseteq \Sigma'$ and $S \in \Sigma^\infty$, $\func{dim}_{FS}^{(\Sigma')}(S)$ treats each character $a \in \Sigma$ in $S$ as a character from $\Sigma'$. We rely on context to distinguish these two scenarios. 

The following theorem establishes the relationship between the finite-state dimension of a binary sequence and its finite-state dimension when viewed as a sequence drawn from $A \subseteq \{0,1\}^l$.

\begin{theorem}
Let $l \in \mathbb{Z}^+$ and $\emptyset \neq A \subseteq \{0,1\}^l$. Then, for all $S \in A^\infty$,
$ \func{dim}_{FS}^{(\{0,1\})}(S) = \frac{\log |A|}{l} \func{dim}_{FS}^{(A)}(S).$
\label{thm-A}
\end{theorem}

\subsection{Variations on the Champernowne sequence}
\label{sec-champ}
This section presents two variations on the Champernowne sequence \cite{champernowne} and shows them to be normal.

First we need the following lemma, which establishes that splicing two normal sequences together results in a normal sequence, as long as the splicing takes increasingly longer substrings from each sequence.

\begin{lemma}
Let $S,T \in \Sigma^\infty$ be normal over the alphabet $\Sigma$. Let
$$U = S[1 \twodots i_1] T[1 \twodots i_1] S[i_1+1 \twodots i_2] T[i_1+1 \twodots i_2] S[i_2+1 \twodots i_3] T[i_2+1 \twodots i_3]  \ldots$$
such that the function $f(j) = i_{j+1} - i_j$ is nondecreasing and unbounded. Then $U$ is normal over the alphabet $\Sigma$.
\label{lem-splice}
\end{lemma}

Let $d \in (0,1) \cap \mathbb{Q}$. Since $d$ is rational, $d = \frac{d_n}{d_d}$ for some $d_n,d_d \in \mathbb{Z}^+$. Since $d<1$, $d_d \geq 2$. Since $d > 0$, $d_n \geq 1$.  Let $l=d_d$ and let $A \subseteq \{0,1\}^l - \{1^l\}$ such that $\log |A|=d_n$. Note that $|A| = 2^{d l}$, and that since $d_n \geq 1$, $|A| \geq 2$.

Let $\alpha_i \in A^*$ be the string consisting of all strings of length $i$ over the alphabet $A$, concatenated in lexicographical ordering. Let $c=1^l$. Define the sequences
\begin{eqnarray*}
C  &=& \alpha_1 \overline{\alpha_1} \alpha_2 \alpha_2 \overline{\alpha_2 \alpha_2} \alpha_3 \alpha_3 \alpha_3 \overline{\alpha_3 \alpha_3 \alpha_3} \ldots \\
C' &=& \alpha_1 c \overline{\alpha_1} \alpha_2 \alpha_2 c \overline{\alpha_2 \alpha_2} \alpha_3 \alpha_3 \alpha_3 c \overline{\alpha_3 \alpha_3 \alpha_3} \ldots
\end{eqnarray*}
Note that $|\alpha_i| = i |A|^i l \implies |\alpha_i^i| = i^2 |A|^i l$. Champernowne \cite{champernowne} and Merkel and Reimann \cite{merkel-reimann} showed that the sequence $\alpha_1 \alpha_2^2 \alpha_3^3 \ldots$ is normal over the alphabet $A$.

\begin{lemma}
Let
$R=\overline{\alpha_1} \ \overline{\alpha_2^2} \ \overline{\alpha_3^3} \ldots.$
Then $R$ is normal over alphabet $\Sigma$.
\label{lem-rev-normal}
\end{lemma}

\begin{lemma}
$C$ is normal over the alphabet $\Sigma$.
\label{lem-C-normal}
\end{lemma}

Note, however, that $C$ and $C'$ are not normal to base 2, because no more than $2l - 2$ 1's appear consecutively in either sequence. In fact, they have dimension equal to $d$, as established by the following lemma.

\begin{lemma}
$\func{dim}_{FS}^{(\{0,1\})}(C') = d.$
\label{lem-dim-fs}
\end{lemma}

\subsection{Pushdown gambling on a marked sequence}

The sequence $C'$ presented in section \ref{sec-champ} has pushdown dimension bounded above by half of its finite state dimension.

\begin{lemma}
$\func{dim}_{PD}^{(\{0,1\})}(C') \leq \frac{1}{2} d$.
\label{lem-dim-pd}
\end{lemma}

The main theorem of the paper follows and establishes that the pushdown dimension of $C'$ is bounded above by half of its finite-state dimension.

\begin{theorem}
$\func{dim}_{PD}^{(\{0,1\})}(C') \leq \frac{1}{2} \func{dim}_{FS}^{(\{0,1\})}(C')$.
\label{thm-main}
\end{theorem}

Recall that $d \in (0,1) \cap \mathbb{Q}$. Therefore Theorem \ref{thm-main} implies that for every rational $0 < d < 1$, there exists a sequence $C'$ with finite-state dimension $d$ such that $\func{dim}_{PD}(C') \leq \frac{1}{2} \func{dim}_{FS}(C')$.

\section{Conclusion}
\label{conclusion}

We have shown that there exist sequences with pushdown dimension strictly less than their finite-state dimension. This was done by the addition of special marker strings that are placed increasingly far apart in the sequence. Because these marker strings do not occur in other parts of the sequence, the sequence is not normal, and this prevents our proof from showing that any normal sequence has pushdown dimension less than 1. The marker strings are needed for our proof, but it is not known whether they are essential to bound the pushdown dimension. It is possible that the original sequence, without the markers, has the same pushdown dimension.

Nichols \cite{nichols} has shown that there is a normal sequence $S$ such that a pushdown gambler can succeed on $S$, whereas the normality of $S$ establishes that no finite-state gambler can succeed on $S$. However, the pushdown gambler fails to show that $\func{dim}_{PD}(S) < 1$, since the gambler makes money so slowly that it fails on $S$ if any money is taken away at each step (i.e. if $s < 1$).

\begin{question}
Is there a normal sequence $S$ such that $\func{dim}_{PD}(S) < 1$?
\end{question}

We have shown that there exist sequences $C'$ such that $\func{dim}_{PD}(C') \leq \frac{1}{2} \func{dim}_{FS}(C')$. The factor $\frac{1}{2}$ seems artificial, and in our proof, it is an artifact of the particular pushdown gambler we designed. It is an open question whether this could be strengthened to show a larger separation between pushdown and finite-state dimension.

\begin{question}
Is there a sequence $S$ such that $\func{dim}_{PD}(S) < \frac{1}{2} \func{dim}_{FS}(S)$?
\end{question}

It is known that the dimension of an individual sequence with respect to some complexity classes can be equivalently expressed in terms of asymptotic compression ratios \cite{fsd} and logarithmic-loss prediction \cite{logloss-pred}. Furthermore, finite-state dimension can be equivalently characterized in terms of entropy rates \cite{ziv-lempel}, and constructive dimension can be characterized in terms of constructive entropy rates \cite{hitchcock-thesis}. It would be useful to find other such alternative characterizations of the resource-bounded dimensions of sequences.

\section{Acknowledgements}
The authors thank Jack Lutz, Jim Lathrop, and Giora Slutzki for their help and for useful discussions.

\newpage

\section{Appendix}


\begin{proof}[Proof of Lemma \ref{lem-add-chars-dim}]
We first show that $\func{dim}_{FS}^{(\Sigma')}(S) \leq \frac{\log |\Sigma|}{\log |\Sigma'|} \func{dim}_{FS}^{(\Sigma)}(S)$.
\begin{description}
    \item[]

Let $s > \func{dim}_{FS}^{(\Sigma)}(S)$. Then there exists an FSG $G=(Q,\Sigma,\delta,\beta,q_0)$ that $s$-succeeds on $S$. Construct the FSG $G'=(Q',\Sigma',\delta',\beta',q'_0)$ as follows
\begin{itemize}
    \item $Q'=Q$,
    \item $\delta'(q,a) = \left\{%
\begin{array}{ll}
   \delta(q,a) , & \text{if $a \in \Sigma$} \\
   \bot , & \text{otherwise} \\
\end{array}%
\right.    $,
    \item $\beta'(q)(a) = \left\{%
\begin{array}{ll}
   \beta(q)(a) , & \text{if $a \in \Sigma$} \\
   0 , & \text{otherwise} \\
\end{array}%
\right.    $,
    \item $q'_0=q_0$.
\end{itemize}

Since $S$ contains no characters from $\Sigma' - \Sigma$, for all $n \in \nat$,
$$
d_{G'}(S_n)
=
\left( \frac{|\Sigma'|} {|\Sigma|} \right)^n d_G(S_n).
$$
Let $t = s \frac{\log |\Sigma|}{\log |\Sigma'|}$. Then
\begin{eqnarray*}
d_{G'}^{(t)}(S_n)
&\triangleq&
|\Sigma'|^{(t-1)n} d_{G'}(S_n)
\\ &=&
|\Sigma'|^{(t-1)n} \left( \frac{|\Sigma'|} {|\Sigma|} \right)^n d_G(S_n)
\\ &=&
\left( \frac{|\Sigma'|^t} {|\Sigma|} \right)^n d_G(S_n)
\\ &=&
\left( \frac{|\Sigma'|^{ s \frac{\log |\Sigma|}{\log |\Sigma'|} }} {|\Sigma|} \right)^n d_G(S_n)
\\ &=&
|\Sigma|^{(s-1)n} d_G(S_n)
\\ &\triangleq&
d_G^{(s)}(S_n).
\end{eqnarray*}

    Thus $G'$ $t$-succeeds on $S$, since $G$ $s$-succeeds on $S$. Since this holds for every $s > \func{dim}_{FS}^{(\Sigma)}(S)$, $\func{dim}_{FS}^{(\Sigma')}(S) \leq \frac{\log |\Sigma|}{\log |\Sigma'|} \func{dim}_{FS}^{(\Sigma)}(S)$.

\item[]
    We next show that $\func{dim}_{FS}^{(\Sigma')}(S) \geq \frac{\log |\Sigma|}{\log |\Sigma'|} \func{dim}_{FS}^{(\Sigma)}(S)$.

    Let $t > \func{dim}_{FS}^{(\Sigma')}(S)$. Then there exists an FSG $G=(Q,\Sigma',\delta,\beta,q_0)$  that $t$-succeeds on $S$. Since $S$ contains no characters from $\Sigma' - \Sigma$, assume without loss of generality that $\beta(q,a)=0$ for all $q \in Q$ and all $a \in \Sigma' - \Sigma$. This assumption can be made for the following reason. If a gambler does bet non-zero capital on $a \in \Sigma' - \Sigma$, we can always construct a gambler that takes the capital $G$ bets on $a$ and uniformly distributes it to the remaining characters in $\Sigma$. Since $a$ does not appear in $S$, this new gambler will make strictly more money than the old, and hence will $s$-succeed whenever the old gambler does.

    Then a straightforward reversal of the previous direction of the proof suffices to show that there is a gambler $G'=(Q',\Sigma,\delta',\beta',q'_0)$ that $s$-succeeds on $S$, where $s = t \frac{\log |\Sigma'|}{\log |\Sigma|}$. This establishes that $\func{dim}_{FS}^{(\Sigma')}(S) \geq \frac{\log |\Sigma|}{\log |\Sigma'|} \func{dim}_{FS}^{(\Sigma)}(S)$.

\end{description}

\end{proof}


\begin{proof}[Proof of Lemma \ref{lem-dim-same}]
Let $S$ and $S'$ be as in the statement of the lemma. Let $l \in\mathbb{Z}^+$, and let $w \in \Sigma^l$.

Let there be $k_n$ insertions of the marker character $m$ in $S_n$ (i.e. the insertion indices satisfy $1 \leq i_1 < i_2 < \ldots < i_{k_n} \leq n < i_{k_n+1}$). Then $S'_{n+k_n}$ is the prefix of $S'$ ``corresponding'' to $S_n$. Note that $freq(m,S'_{n+k_n}) = \frac{k_n}{n+k_n}$.

Since $i_{j+1}-i_j$ is non-decreasing and unbounded, $(\forall p \in \nat)(\exists n_p)$ such that all markers after position $n_p$ are at least $p$ characters apart. Hence $freq(m,S'[n_p \twodots]) \leq \frac{1}{p}$. By Lemma \ref{lem-prefix}, $freq(m,S') \leq \frac{1}{p}$. Since this holds for all $p \in \nat$, $freq(m,S') = 0$. Since $freq(m,S'_{n+k_n}) = \frac{k_n}{n+k_n}$, then $k_n = o(n)$; $k_n$ grows strictly slower than $n$.

Since there are $k_n$ occurrences of $m$ in $S'_{n+k_n}$, there are $k_n (l-1)$ substrings of length $l$ in $S_n$ that could have been changed by having an $m$ inserted into them. In the worst case, every one of these substrings was our chosen string $w$. Thus
\begin{equation}
\underbrace{\#(w,S'_{n+k_n})}_{\begin{array}{l} \# \text{ of $w$ in $S'_{n+k_n}$} \end{array}}
\geq \underbrace{\#(w,S_n)}_{\begin{array}{l} \# \text{ of $w$ in $S_n$} \end{array}}
- \underbrace{k_n (l-1)}_{\begin{array}{l} \# \text{ of $w$ in $S_n$ that} \\ \text{could have changed} \end{array} }
\label{geq-num}
\end{equation}

Since $w \in \Sigma^l$, it does not contain an $m$. Adding $m$'s to $S$ cannot add more $w$'s to $S$. Thus
\begin{equation}
\#(w,S'_{n+k_n}) \leq \#(w,S_n)
\label{leq-num}
\end{equation}
Recall that $k_n = o(n)$. Thus
\begin{eqnarray*}
\lim_{n \toinf} \left( freq(w,S'_{n+k_n}) - freq(w,S_n) \right)
&=&
\lim_{n \toinf} \left( \frac{\#(w,S'_{n+k_n})}{n+k_n-l+1} - \frac{\#(w,S_n)}{n-l+1} \right)
\\ &\geq&
\lim_{n \toinf} \left( \frac{\#(w,S_n) - k_n (l-1)}{n+k_n-l+1} - \frac{\#(w,S_n)}{n-l+1} \right) \ \ \ \ \text{inequality } \eref{geq-num}
\\ &=&
\lim_{n \toinf} \left( \frac{\#(w,S_n) - k_n (l-1)}{n-l+1} - \frac{\#(w,S_n)}{n-l+1} \right) \ \ \ \ \text{since } k_n = o(n)
\\ &=&
\lim_{n \toinf} \left( \frac{- k_n (l-1)}{n-l+1} \right) 
\\ &=&
0, \ \ \ \ \text{since } k_n = o(n)
\end{eqnarray*}
and
\begin{eqnarray*}
\lim_{n \toinf} \left( freq(w,S'_{n+k_n}) - freq(w,S_n) \right)
&=&
\lim_{n \toinf} \left( \frac{\#(w,S'_{n+k_n})}{n+k_n-l+1} - \frac{\#(w,S_n)}{n-l+1} \right)
\\ &\leq&
\lim_{n \toinf} \left( \frac{\#(w,S_n) }{n+k_n-l+1} - \frac{\#(w,S_n)}{n-l+1} \right)  \ \ \ \ \text{inequality } \eref{leq-num}
\\ &=&
\lim_{n \toinf} \left( \frac{\#(w,S_n) }{n-l+1} - \frac{\#(w,S_n)}{n-l+1} \right) \ \ \ \ \text{since } k_n = o(n)
\\ &=&
0.
\end{eqnarray*}
Thus
$$
\lim_{n \toinf} \left( freq(w,S'_{n+k_n}) - freq(w,S_n) \right) = 0
$$
This establishes that $freq(w,S_n)$ and $freq(w,S'_{n+k_n})$ approach each other as $n \toinf$, for all $w \in \Sigma^l$. Let $w' \in \Sigma_m^l - \Sigma^l$. Then $freq(w',S_n)=0$ for all $n$, since no $m$'s appear in $S$. Since $freq(m,S')=0$,
\begin{eqnarray*}
freq(w',S')
&\triangleq&
\lim_{n \toinf} \frac{\#(w',S'_n)}{n}
\\ &\leq&
\lim_{n \toinf} \frac{l \#(m,S'_n)}{n}
\\ &=&
l \cdot freq(m,S')
\\ &=&
0,
\end{eqnarray*}
where the inequality follows from the fact that for each $m$ that appears in $S'_n$, at most $l$ substrings of length $l$ in $S'_n$ could have that $m$ in them, and hence belong to $\Sigma_m^l - \Sigma^l$.
By the non-negativity of $freq, freq(w',S') = 0 = freq(w',S)$, implying
$$\lim\limits_{n \toinf} \left( freq(w',S_n) - freq(w',S'_{n+k_n}) \right) = 0$$
for all $w' \in \Sigma_m^l - \Sigma^l$. Hence,
\begin{equation}
\left( \forall w \in \Sigma_m^l \right) \ \lim_{n \toinf} \left( freq(w,S'_{n+k_n}) - freq(w,S_n) \right) = 0
\label{eq-lim-zero}
\end{equation}
Thus
\begin{eqnarray*}
H_l(S')
&\triangleq&
\frac{1}{l \log |\Sigma_m|} \liminf_{n \toinf} \sum_{w \in \Sigma_m^l} freq(w,S'_n) \log \frac{1}{freq(w,S'_n)}
\\ &=&
\frac{1}{l \log |\Sigma_m|} \liminf_{n \toinf} \sum_{w \in \Sigma_m^l} freq(w,S'_{n+k_n}) \log \frac{1}{freq(w,S'_{n+k_n})}
\\ &=&
\frac{1}{l \log |\Sigma_m|} \liminf_{n \toinf} \sum_{w \in \Sigma_m^l} freq(w,S_n) \log \frac{1}{freq(w,S_n)} \ \ \ \ \text{by } \eref{eq-lim-zero}
\\ &\triangleq&
H_l(S).
\end{eqnarray*}

Since this holds for all $l$, $H(S) = H(S')$. By Lemma \ref{lem-dim-h-eq}, $\func{dim}_{FS}^{(\Sigma_m)}(S) = \func{dim}_{FS}^{(\Sigma_m)}(S')$.

\end{proof}


\begin{proof}[Proof of Theorem \ref{thm-A}]
\hfill
\begin{description}

    \item[]
We first show that $\func{dim}_{FS}^{(\{0,1\})}(S) \geq \frac{\log |A|}{l} \func{dim}_{FS}^{(A)}(S)$.

This holds trivially if $|A|=1$, so assume $|A| \geq 2$. Let $s \in [0,\infty) \cap \mathbb{Q}$ such that $s > \func{dim}_{FS}(S)$.

By our choice of $s$, there exists an FSG $G=(Q,\{0,1\},\delta,\beta,q_0)$ such that $G$ $s-$succeeds on $S$. Construct and FSG $G'=(Q',\Sigma',\delta',\beta',q'_0)$ as follows:

    \begin{itemize}
        \item $Q'=Q$.
        \item $\Sigma'=A$.
        \item for all $q \in Q'$ and $w \in A$,
        $$\delta'(q,w)=\delta(q,w).$$
        \item for all $q \in Q'$ and $w \in A$,
        $$\beta'(q)(w) = \left\{%
\begin{array}{ll}
    \frac{ \widetilde{B}(q)(w) }{ \widetilde{B}(q)(A) }, & \hbox{if $\widetilde{B}(q)(A)>0$;} \\
    0, & \hbox{if $\widetilde{B}(q)(A)=0$.} \\
\end{array}%
\right.   , $$
where
$$
\widetilde{B}(q)(w) = \prod_{i=1}^l \beta(\delta(q,w_{i-1}))(w[i])
$$
and
$$
\widetilde{B}(q)(A) = \sum_{w \in A} \widetilde{B}(q)(w).
$$
        \item $q'_0=q_0$.
\end{itemize}

Note that for all $q \in Q'$, $d_{G',q}$ is a martingale, and that $A \subseteq \{0,1\}^l$ is a prefix set. Let $q \in Q'$. Then
\begin{eqnarray*}
\widetilde{B}(q)(A) &\triangleq& \sum_{w \in A} \widetilde{B}(q)(w)
\\ &=&
\sum_{w \in A} \prod_{i=1}^l \beta(\delta(q,w_{i-1}))(w[i])
\\ &=&
\sum_{w \in A} d_{G',q}(w)
\\ &\leq&
1. \ \ \ \ \text{by corollary \ref{cor-kraft}}
\end{eqnarray*}

So
\begin{equation}
\widetilde{B}(q)(A) \leq 1
\label{bqa-leq-1}
\end{equation}

for all $q \in Q'$.

Let $w \in A$ and let $q \in Q'$. Then
\begin{eqnarray*}
d_{G',q}(w) &=& \frac{ \widetilde{B}(q)(w) }{ \widetilde{B}(q)(A) }
\\&\geq&
\widetilde{B}(q)(w) \ \ \ \ \ \ \ \ \text{by \eref{bqa-leq-1}}
\\&=&
\prod_{i=1}^l \beta(\delta(q,w_{i-1}))(w[i])
\\&=&
d_{G,q}(w).
\end{eqnarray*}

So by induction, for all $z \in A^*$,
\begin{equation*}
d_G'(z) \geq d_G(z).
\label{dg'-leq-dg}
\end{equation*}

Let $z \in A^*$ and $w \in A$, and let $q = \delta(z)$. Then
\begin{eqnarray*}
d_G(zw)
&=&
2^l \widetilde{B}(q)(w) d_G(z)
\\ &\leq&
2^l \widetilde{B}(q)(w) d_{G'}(z)
\\ \Rightarrow
d_{G'}(z)
&\geq&
\frac{1}{2^l \widetilde{B}(q)(w)} d_G(zw)
\end{eqnarray*}
and so
\begin{eqnarray*}
d_{G'}(zw)
&=&
|A| \frac{ \widetilde{B}(q)(w) }{ \widetilde{B}(q)(A) } d_{G'}(z)
\\ &\geq&
|A| \frac{ \widetilde{B}(q)(w) }{ \widetilde{B}(q)(A) } \frac{1}{2^l \widetilde{B}(q)(w)} d_G(zw)
\\ &=&
\frac{|A|}{2^l \widetilde{B}(q)(A)} d_G(zw)
\\ &\geq&
\frac{|A|}{2^l} d_G(zw). \ \ \ \ \text{inequality \eref{bqa-leq-1}}
\end{eqnarray*}
So by induction
\begin{equation}
d_{G'}(z) \geq \left( \frac{|A|}{2^l} \right)^{\frac{|z|}{l}} d_G(z).
\label{dg'-leq-dg-extra}
\end{equation}

Let $t = \frac{sl}{\log |A|}$. Then
\begin{eqnarray*}
d_{G'}^{(t)}(z)
&\triangleq&
|A|^{(t-1) \frac{|z|}{l}} d_{G'}(z)
\\ &\geq&
|A|^{(t-1) \frac{|z|}{l}} \left( \frac{|A|}{2^l} \right)^{\frac{|z|}{l}} d_G(z) \ \ \ \ \text{inequality \eref{dg'-leq-dg-extra}}
\\ &=&
\frac{ |A|^{\frac{t |z|}{l}} }{ 2^{|z|} } d_G(z)
\\ &=&
\frac{ |A|^{\frac{s |z|}{ \log |A| }} }{ 2^{|z|} } d_G(z)
\\ &\geq&
\frac{ 2^{\frac{s |z|}{ \log |A| }} }{ 2^{|z|} } d_G(z) \ \ \ \ \text{since $|A| \geq 2$}
\\ &\geq&
\frac{ 2^{s |z|} }{ 2^{|z|} } d_G(z)  \ \ \ \ \text{since $|A| \geq 2$}
\\ &=&
2^{(s-1) |z|} d_G(z)
\\ &\triangleq&
d_G^{(s)}(z).
\end{eqnarray*}
Thus $G'$ $t$-succeeds whenever $G$ $s$-succeeds. This establishes that
$$\func{dim}_{FS}^{(\{0,1\})}(S) \geq \frac{s}{t} \func{dim}_{FS}^{(A)}(S) = \frac{\log |A|}{l} \func{dim}_{FS}^{(A)}(S).$$

    \item[]
We next show that $\func{dim}_{FS}^{(\{0,1\})}(S) \leq \frac{\log |A|}{l} \func{dim}_{FS}^{(A)}(S)$.

Let $s \in [0,\infty) \cap \mathbb{Q}$ such that $s > \func{dim}_{FS}^{(A)}(S)$, and let $t = \frac{s \log |A|}{l}$. Then it suffices to show that $\func{dim}_{FS}^{(\{0,1\})}(S) \leq t$. By our choice of $s$, there exists an FSG $G=(Q,A,\delta,\beta,q_0)$ such that $G$ $s$-succeeds on $S$.

Let $ppref(A)$ be the set of all proper prefixes of the strings in $A$. Construct the FSG $G'=(Q',\Sigma',\delta',\beta',q'_0)$ as follows:
\begin{itemize}
    \item $Q' = Q \times ppref(A)$.
    \item $\Sigma' = \{0,1\}$.
    \item for all $q \in Q'$, $w \in ppref(A)$, and $b \in \{0,1\}$.
    \begin{eqnarray*}
    \delta'((q,w),b)
    &=&
    \left\{%
\begin{array}{ll}
    (q,wb), & \hbox{if $wb \in ppref(A)$;} \\
    (\delta(q,wb),\lambda), & \hbox{if $wb \in A$;} \\
    \bot, & \hbox{otherwise.} \\
   \end{array}%
\right. .
\end{eqnarray*}
    \item for all $q \in Q'$, $w \in ppref(A)$, and $b \in \{0,1\}$,
$$
    \beta'(q,w)(b) = \left\{%
\begin{array}{ll}
    \frac{ \widetilde{B}(q,wb) }{ \widetilde{B}(q,w) }, & \hbox{if $\widetilde{B}(q,w)>0$;} \\
    0, & \hbox{if $\widetilde{B}(q,w)=0$.} \\
   \end{array}%
\right. ,
$$
where
$$
\widetilde{B}(q,w) = \sum_{u \in A(w)} \beta(q)(wu)
$$
and
$$
A(w) = \{ u \in \{0,1\}^* \ |\ wu \in A \}.
$$
    \item $q'_0 = (q_0,\lambda)$.
   \end{itemize}

In the non-degenerate case (where $\widetilde{B}(q,w) > 0$)
$$
\beta'(q,w)(0) + \beta'(q,w)(1) = \frac{ \widetilde{B}(q,w0) + \widetilde{B}(q,w1) }{ \widetilde{B}(q,w) }.
$$
For all $w \in ppref(A)$, $A(w)$ is the disjoint union of $A(w0)$ and $A(w1)$. So $\widetilde{B}(q,w0) + \widetilde{B}(q,w1) = \widetilde{B}(q,w)$. Therefore $\beta'(q,w)(0) + \beta'(q,w)(1) = 1$.

Note that for all $q \in Q$, $\widetilde{B}(q,w) \leq 1$ for all $w \in ppref(A) \cup A$. This follows from the fact that $w = \lambda$ maximizes $\widetilde{B}(q,w)$. $\widetilde{B}(q,\lambda) = \sum\limits_{w \in A} \beta(q)(w) = 1$, by the constraint that $\beta(q)$ is a probability measure over $A$.

Intuitively, $G'$'s martingale bets $l$ times every $l$ bits, in such a way that the $l$ bets made will telescope to simulate the bet made once every $l$ bits by $G$.

Let $z \in A^*$, $w \in A$, and $q = \delta(z)$. Then
\begin{eqnarray*}
d_{G'}(zw)
&=&
2^l d_{G'}(z) \prod_{i=1}^l \beta'(q,w_{i-1})(w[i])
\\ &=&
2^l d_{G'}(z) \prod_{i=1}^l \frac{ \widetilde{B}(q,w_i) }{ \widetilde{B}(q,w_{i-1})}
\\ &=&
2^l d_{G'}(z) \frac{ \widetilde{B}(q,w) }{ \widetilde{B}(q,\lambda)}
\\ &\geq&
2^l d_{G'}(z) \widetilde{B}(q,w)
\\ &=&
2^l d_{G'}(z) \beta(q)(w),
\end{eqnarray*}
and
$$
d_G(zw) = |A| \beta(q)(w) d_G(z).
$$
So by induction
$$
d_{G'}(z) \geq \prod_{i=1}^{\frac{|z|}{l}} 2^l \beta(\delta(z_{il}))(w),
$$
and
$$
d_G(z) = \prod_{i=1}^{\frac{|z|}{l}} |A| \beta(\delta(z_{il}))(w).
$$
So
\begin{eqnarray*}
\frac{d_{G'}(z)}{d_G(z)}
&\geq&
\frac{ \prod\limits_{i=1}^{\frac{|z|}{l}} 2^l \beta(\delta(z_{il}))(w)  }{ \prod\limits_{i=1}^{\frac{|z|}{l}} |A| \beta(\delta(z_{il}))(w) }
\\ &=&
\left( \frac{2^l}{|A|} \right)^{\frac{|z|}{l}}
\\ \Rightarrow
d_{G'}(z)
&\geq&
\left( \frac{2^l}{|A|} \right)^{\frac{|z|}{l}} d_G(z).
\end{eqnarray*}
Thus
\begin{eqnarray*}
d_{G'}^{(t)}(z)
&\triangleq&
2^{(t-1) |z|} d_{G'}(z)
\\ &\geq&
2^{(t-1) |z|} \left( \frac{2^l}{|A|} \right)^{\frac{|z|}{l}} d_G(z)
\\ &=&
2^{t |z|} |A|^{- \frac{|z|}{l}} d_G(z)
\\ &=&
2^{t |z| - \frac{|z|}{l} \log |A|} d_G(z)
\\ &=&
2^{\frac{s \log |A|}{l} |z| - \frac{|z|}{l} \log |A|} d_G(z)
\\ &=&
|A|^{(s-1) \frac{|z|}{l}} d_G(z)
\\ &\triangleq&
d_G^{(s)}(z).
\end{eqnarray*}
Therefore $G'$ $t$-succeeds when $G$ $s$-succeeds.
This establishes that
$$\func{dim}_{FS}^{(\{0,1\})}(S) \leq \frac{t}{s} \func{dim}_{FS}^{(A)}(S) = \frac{\log |A|}{l} \func{dim}_{FS}^{(A)}(S).$$

\end{description}

\end{proof}


\begin{proof}[Proof of Lemma \ref{lem-splice}]
Let $n = i_j$, for some $j \in \mathbb{Z}^+$. Let $k_n = j$. Intuitively, $k_n$ is the number of splices taken from $S_n$ and $T_n$ apiece to form $z_{2n}$. Since $i_{j+1} - i_j$ is nondecreasing and unbounded, $\lim\limits_{n \toinf} \frac{k_n}{n} = 0$.

Let $l \in \mathbb{Z}^+$, and let $w \in \Sigma^l$. Then $freq(w,S)=freq(w,T)=|\Sigma|^{-l}$. Because there are only $k_n$ places in $S_n$ at which it was ``broken'' to be spliced into $T_n$, at most $k_n (l-1)$ instances of $w$ in $S_n$ could have been disrupted by the splicing and hence not appear in $z_{2n}$. The same argument applies to instances of $w$ in $T_n$. Thus
$$
\#(w,z_{2n})
\geq
\#(w,S_n) + \#(w,T_n) - 2 k_n (l-1)
$$
Therefore
\begin{eqnarray*}
freq(w,z)
&\triangleq&
\lim\limits_{n \toinf} \frac{\#(w,z_n)}{n}
\\ &=&
\lim\limits_{n \toinf} \frac{\#(w,z_{2n})}{2n}
\\ &\geq&
\lim\limits_{n \toinf} \frac{\#(w,S_n) + \#(w,T_n) - 2 k_n (l-1)}{2n}
\\ &=&
\frac{1}{2} \lim\limits_{n \toinf} \frac{\#(w,S_n)}{n} + \frac{1}{2} \lim\limits_{n \toinf} \frac{\#(w,T_n)}{n} - (l-1) \lim\limits_{n \toinf} \frac{k_n}{n}
\\ &=&
\frac{1}{2} \lim\limits_{n \toinf} \frac{\#(w,S_n)}{n} + \frac{1}{2} \lim\limits_{n \toinf} \frac{\#(w,T_n)}{n}
\\ &\triangleq&
\frac{1}{2} freq(w,S) + \frac{1}{2} freq(w,T)
\\ &=&
|\Sigma|^{-l}.
\end{eqnarray*}

This holds for all $w \in \Sigma^l$. $\sum\limits_{w \in \Sigma^l} freq(w,z) = 1$, so, for all $w \in \Sigma^l$, $freq(w,z) = |\Sigma|^{-l}$.

Since this holds for all $l \in \pint$,  $z$ is normal.

\end{proof}


\begin{proof}[Proof of Lemma \ref{lem-rev-normal}]
Let $R(k) = \overline{\alpha_1} \ \overline{\alpha_2^2} \ \overline{\alpha_3^3} \ldots \overline{\alpha_k^k}$. Let $n_k = |R(k)|$, so that $R_{n_k} = R(k)$. Let $U = \alpha_1 \alpha_2^2 \alpha_3^3 \ldots$.

Let $l \in \mathbb{Z}^+$, and let $w \in \Sigma^l$. Then $freq(w,U)=|\Sigma|^{-l}$, by the normality of $U$. For every instance of $w$ in $x$ that does not cross a boundary between $\alpha_i^i$ and $\alpha_{i+1}^{i+1}$, then a corresponding instance of $\overline{w}$ appears in $R$, since $R=\overline{\alpha_1} \ \overline{\alpha_2^2} \ \overline{\alpha_3^3} \ldots$. There are at most $(k-1)(l-1)$ instances of $w$ that could lie across such a boundary. Since $f(k) = n_{k+1} - n_k$ is nondecreasing and unbounded, $\lim\limits_{k \toinf} \frac{k}{n_k}=0$.

Therefore, for all $k \in \pint$ and all $w \in \Sigma^l$
$$
\#(\overline{w},R_{n_k})
\geq
\#(w,U_{n_k}) - (k-1)(l-1)
$$
So
\begin{eqnarray*}
freq(\overline{w},R)
&\triangleq&
\lim\limits_{n \toinf} \frac{\#(\overline{w},R_n)}{n}
\\ &=&
\lim\limits_{k \toinf} \frac{\#(\overline{w},R_{n_k})}{n_k}
\\ &\geq&
\lim\limits_{k \toinf} \frac{\#(w,U_{n_k}) - (k-1)(l-1)}{n_k}
\\ &=&
\lim\limits_{k \toinf} \frac{\#(w,U_{n_k})}{n_k} - \lim\limits_{k \toinf} \frac{(k-1)(l-1)}{n_k}
\\ &=&
\lim\limits_{k \toinf} \frac{\#(w,U_{n_k})}{n_k}
\\ &=&
\lim\limits_{n \toinf} \frac{\#(w,U_n)}{n}
\\ &\triangleq&
freq(w,U)
\\ &=&
\Sigma^{-l}
\end{eqnarray*}

Since this holds for the reverse of every string in $\Sigma^l$, it holds for all $w \in \Sigma^l$. Since $$\sum\limits_{w \in \Sigma^l} freq(w,R) = 1,$$ $freq(w,R) = |\Sigma|^{-l}$, for all $w \in \Sigma^l$.

Since this holds for all $l \in \pint$, $R$ is normal.

\end{proof}


\begin{proof}[Proof of Lemma \ref{lem-C-normal}]
This follows immediately from Lemmas \ref{lem-rev-normal} and \ref{lem-splice} and the normality of $\alpha_1 \alpha_2^2 \alpha_3^3 \ldots$.

\end{proof}


\begin{proof}[Proof of Lemma \ref{lem-dim-pd}]
Let $s > s' > d$. It suffices to show that $\func{dim}_{PD}^{(\{0,1\})}(C') \leq \frac{1}{2} s$. Let $A_c = A \cup \{ c \}$.


We construct a PDG $P$ that does the following. It reads the sequence $C' = \alpha_1 c \overline{\alpha_1} \alpha_2^2 c \overline{\alpha_2^2} \ldots $ in two alternating stages. The first stage involves reading the substring $\alpha_i^i c$, and the second stage involves reading the substring $\overline{\alpha_i^i}$. In the first stage, $P$ bets optimally for any FSG, while the bits it reads are pushed onto the stack. Once $c$ has been read, $P$ pops $c$ from the stack, and then uses the string it pushed, which is $\alpha_i^i$, to bet optimally on the string that follows, which is $\overline{\alpha_i^i}$. It pops bits until the stack is empty, at which point $\alpha_{i+1}^{i+1}$ follows, and the gambler begins again.



As $P$ is pushing bits onto its stack, it bets an equal amount ($d_P(a) = 2^l \frac{1 - \epsilon}{|A|}$) on all bitstrings $a \in A$. It bets a small amount ($d_P(c) = 2^l \epsilon$) on the bitstring $c=1^l$, and this bet can be made vanishingly smaller by shrinking $\epsilon$, although some positive bet must be made so $P$'s capital does not become 0 when it encounters $c$. The requirement that $\epsilon < 1 - 2^{l(s' - s)}$ ensures that $P$ $\left( \frac{1}{2} s \right)$-succeeds on $C'$, which is shown formally below. $P$ bets nothing on any bitstring $a \not\in A_c$.



Thus, $P$'s martingale behaves optimally for any finite-state martingale when reading the subsequence $\alpha_1 c \alpha_2^2 c \ldots \alpha_i^i c \ldots$, and it doubles its money on every bit when reading the subsequence $\overline{\alpha_1} \overline{\alpha_2^2} \ldots \overline{\alpha_i^i} \ldots$.

Formally, the PDG $P=(Q',\Sigma',\Gamma',\delta',\beta',q'_0,z)$ is defined as follows on input $S \in \can$:

\begin{codebox}
\Procname{$P(C')$}
\li $i \gets 1$ \>\>\>\>\>\>\>\>\Comment{current bit of $C'$}
\li \While \const{true} \>\>\>\>\>\>\>\>\Comment{each iteration $k$ reads $\alpha_k^k c \overline{\alpha_k^k}$} \li \Do
        \Repeat         \>\>\>\>\Comment{push bits until marker found}
\li     $w \gets \lambda$
\li     \For $j \gets 1$ \To $l$ \>\>\>\>\Comment{set $w$ to next block of length $l$} \li \Do
            bet according to $\beta(w)$ on $C'[i]$
\li         $w \gets w x[i]$
\li         push $C'[i]$ onto stack
\li         $i \gets i + 1$
        \End
\li     \Until $w = 1^l$
\li     pop $l$ bits from stack
\li     \While stack is not empty \li \Do
            bet all capital on bit on top of stack
\li         read $C'[i]$
\li         $i \gets i + 1$
\li         pop 1 bit from stack
        \End
    \End
\end{codebox}

where
\begin{eqnarray*}
    \beta(w)(b) &=& \left\{%
\begin{array}{ll}
    \frac{ \widetilde{B}(wb) }{ \widetilde{B}(w) } , & \hbox{if $\widetilde{B}(w)>0$;} \\
    0, & \hbox{otherwise.} \\
   \end{array}%
\right.
\\
\widetilde{B}(w) &=& \sum_{u \in A_c(w)} B(wu)
\\
A_c(w) &=& \{ u \in \{0,1\}^* \ |\ wu \in A_c \}
\\
B(a) &=& \left\{%
\begin{array}{ll}
    \frac{1 - \epsilon}{|A|}, & \hbox{if $a \in A$;} \\
    \epsilon, & \hbox{if $a = c$;} \\
   \end{array}%
\right.
\\
0 < \epsilon &<& 1 - 2^{l(s' - s)}.
\end{eqnarray*}

Note that, for all $a \in A_c$,
\begin{eqnarray*}
d_P(a)
&=&
2^l \prod_{i=1}^l \beta(a_{i-1})(a[i])
\\ &=&
2^l \prod_{i=1}^l \left( \frac{ \widetilde{B}(a_i) }{ \widetilde{B}(a_{i-1}) } \right)
\\ &=&
2^l \frac{ \widetilde{B}(a) }{ \widetilde{B}(\lambda) }
\\ &=&
2^l \frac{ \sum\limits_{u \in A_c(a)} B(au) }{ \sum\limits_{u \in A_c(\lambda)} B(\lambda u) }
\\ &=&
2^l \frac{ B(a) }{ \sum\limits_{u \in A_c} B(u) }
\\ &=&
2^l \frac{ B(a) }{ \epsilon + \sum\limits_{u \in A} \frac{1 - \epsilon}{|A|} }
\\ &=&
2^l B(a).
\end{eqnarray*}
Thus, for all $a \in A$,
$$
d_P(a) = 2^l \frac{1 - \epsilon}{|A|},
$$
and, for $c = 1^l$
$$
d_P(c) = 2^l \epsilon.
$$


Recall that $d_P(c) = 2^l \epsilon$, and that $P$ makes the same capital ($d_P(a) = 2^l \frac{1 - \epsilon}{|A|}$) on each ``character'' $a \in A$.
\begin{eqnarray*}
d_P(\alpha_k^k)
&=&
\left( d_P(\alpha_k) \right)^k
\\ &=&
\left( d_P(a)^{ \underbrace{|A|^k}_{\text{\# of strings}} \cdot \underbrace{k}_{\text{\# of characters per string}}  }   \right)^k
\\ &=&
\left( 2^l \frac{1 - \epsilon}{|A|} \right)^{ |A|^k k^2 }
\\ &=&
2^{k^2 |A|^k l} \left( \frac{1 - \epsilon}{|A|} \right)^{ k^2 |A|^k },
\end{eqnarray*}
and
\begin{eqnarray*}
d_P(\overline{\alpha_k^k})
&=&
2^{|\overline{\alpha_i^i}|}
\\ &=&
2^{k^2 |A|^k l}.
\end{eqnarray*}
Thus,
\begin{eqnarray*}
d_P(\alpha_k^k c \overline{\alpha_k^k})
&=&
\left( 2^{k^2 |A|^k l} \left( \frac{1 - \epsilon}{|A|} \right)^{ k^2 |A|^k } \right) \left( 2^l \epsilon \right) \left( 2^{k^2 |A|^k} \right)
\\ &=&
\epsilon 2^{2 k^2 |A|^k l + l} \left( \frac{1 - \epsilon}{|A|} \right)^{ k^2 |A|^k }.
\end{eqnarray*}
Let $t = \frac{1}{2} s$. Then
\begin{eqnarray*}
d_G^{(t)}(\alpha_k^k c \overline{\alpha_k^k})
&=&
2^{(t-1) |\alpha_k^k c \overline{\alpha_k^k}|} \epsilon 2^{2 k^2 |A|^k l + l} \left( \frac{1 - \epsilon}{|A|} \right)^{ k^2 |A|^k }
\\ &=&
2^{(t-1) (2 k^2 |A|^k l + l)} \epsilon 2^{2 k^2 |A|^k l + l} \left( \frac{1 - \epsilon}{|A|} \right)^{ k^2 |A|^k }
\\ &=&
\epsilon 2^{t (2 k^2 |A|^k l + l)} \left( \frac{1 - \epsilon}{|A|} \right)^{ k^2 |A|^k }
\\ &=&
2^{tl} \epsilon 2^{t 2 k^2 |A|^k l} \left( \frac{1 - \epsilon}{|A|} \right)^{ k^2 |A|^k }
\\ &=&
2^{tl} \epsilon 2^{t 2 k^2 |A|^k l} \left( \frac{(1 - \epsilon)^{\frac{1}{tl}} }{ |A|^{\frac{1}{tl}} } \right)^{ t l k^2 |A|^k }
\\ &=&
2^{tl} \epsilon \left( 2^2 \frac{(1 - \epsilon)^{\frac{1}{tl}} }{ |A|^{\frac{1}{tl}} } \right)^{ t l k^2 |A|^k }
\\ &=&
2^{tl} \epsilon \left( 2^2 \frac{(1 - \epsilon)^{\frac{1}{tl}} }{ (2^{d l})^{\frac{1}{tl}} } \right)^{ t l k^2 |A|^k }
\\ &>&
2^{tl} \epsilon \left( 2^2 \frac{(1 - \epsilon)^{\frac{1}{tl}} }{ (2^{s' l})^{\frac{1}{tl}} } \right)^{ t l k^2 |A|^k }
\\ &=&
2^{tl} \epsilon \left( 2^2 \frac{(1 - \epsilon)^{\frac{1}{tl}} }{ 2^{\frac{s'}{t}} } \right)^{ t l k^2 |A|^k }
\\ &=&
2^{tl} \epsilon \left( 2^{2 - \frac{s'}{t}} (1 - \epsilon)^{\frac{1}{tl}} \right)^{ t l k^2 |A|^k }
\\ &=&
2^{tl} \epsilon \left( 2^{2 - \frac{2 s'}{s}} (1 - \epsilon)^{\frac{1}{tl}} \right)^{ t l k^2 |A|^k }
\\ &=&
2^{tl} \epsilon \left( 4^{1 - \frac{s'}{s}} (1 - \epsilon)^{\frac{2}{sl}} \right)^{ t l k^2 |A|^k }.
\end{eqnarray*}
Recall that $\epsilon < 1 - 2^{l(s' - s)}$. Then the term in the parentheses,
\begin{eqnarray*}
4^{1 - \frac{s'}{s}} (1 - \epsilon)^{\frac{2}{sl}}
&>&
4^{1 - \frac{s'}{s}} \left( 1 - \left( 1 - 2^{l(s' - s)} \right) \right)^{\frac{2}{sl}}
\\ &=&
4^{1 - \frac{s'}{s}} \left( 2^{l(s' - s)} \right)^{\frac{2}{sl}}
\\ &=&
4^{1 - \frac{s'}{s}} 4^{\frac{1}{s} (s' - s)}
\\ &=&
1.
\end{eqnarray*}

Thus $d_G^{(t)}(\alpha_k^k c \overline{\alpha_k^k})$ grows without bound as $k \toinf$, whence $P$ $t$-succeeds on $C'$. Therefore $$\func{dim}_{PD}^{(\{0,1\})}(C') \leq \frac{1}{2} s \implies \func{dim}_{PD}^{(\{0,1\})}(C') \leq \frac{1}{2} d.$$

\end{proof}


\begin{thebibliography}{99}
{\footnotesize


\bibitem{champernowne} D. G. Champernowne, \emph{The Construction of Decimals Normal in the Scale of Ten}. J. London Math. Soc. 8, 1933.

\bibitem{fsd} J. J. Dai, J. I. Lathrop, J. H. Lutz, and E. Mayordomo. \emph{Finite-state dimension}. Theoretical Computer Science, 310(1-3):1-33, 2004.

\bibitem{hausdorff-book} K. Falconer. \emph{The Geometry of Fractal Sets}. Cambridge University Press, 1985.

\bibitem{feder} M. Feder. \emph{Gambling using a finite-state machine}, IEEE Transcations on Information Theory, 37:1459-1461, 1991.

\bibitem{hausdorff} F. Hausdorff. \emph{Dimension und $\ddot{a}$usseres Mass}. Math. Ann. 79, 157-179, 1919.

\bibitem{logloss-pred}  J. M. Hitchcock. \emph{Fractal dimension and logarithmic loss unpredictability}. Theoretical Computer Science, 2003.

\bibitem{hitchcock-thesis}  J. M. Hitchcock. Effective Fractal Dimension: Foundations and Applications. Ph.D. Dissertation, Iowa State University, 2003.

\bibitem{kozen} D. C. Kozen. \underline{Automata and Computability}. Springer-Verlag New York, Inc., 1997.

\bibitem{kuich-book} W. Kuich, A. Salomaa, \underline{Semirings, automata, languages}, Eatcs Monographs On Theoretical Computer Science; Vol. 5, 1985.

\bibitem{lutz-hausdorff} J. H. Lutz. \emph{Dimension in complexity classes}. SIAM Journal on Computing, 32(5):1236-1259, 2000.

\bibitem{lutz-dim-indv} J. H. Lutz. \emph{The dimensions of individual strings and sequences}. Information and Computation, 187(1):49-79, 2003.

\bibitem{merkel-reimann} W. Merkel and J. Reimann.  \emph{On selection functions that do not preserve normality}.  Theory of Computing Systems.  http://math.uni-heidelberg.de/logic/merkle/ps/normal-mfcsplus.ps

\bibitem{nichols} J. Nichols, \emph{Pushdown gamblers and pushdown dimension}, Master's thesis, Iowa State University, 2004.

\bibitem{schnorr-stimm} C. Schnorr, H. Stimm. \emph{Endliche Automaten und Zufallsfolgen}, Acta Inf. 1: 345-359 1972.

\bibitem{staiger-weighted} L. Staiger, \emph{Weighted Finite Automata and Metrics in Cantor Space}, Journal of Automata, Languages and Combinatorics, vol. 8, no. 2, pg. 353-360, 2003.


\bibitem{ziv-lempel} J. Ziv and A. Lempel. \emph{Compression of individual sequences via variable-rate coding}. IEEE Transactions on Information Theory, 24(5): 530--536, 1978.

}

\end{thebibliography}
\end{document}